\documentclass{emulateapj}
\usepackage{apjfonts}
\usepackage{lscape}
\input{epsf}

\slugcomment{Submitted to the Astrophysical Journal}

\shorttitle{Astrometric companion to the metal-poor star LHS 1589.}
\shortauthors{S. L\'epine {\it et al.}}

\begin{document}

\title{An astrometric companion to the nearby metal-poor,
  low-mass star LHS 1589.\altaffilmark{1,2}}

\author{S\'ebastien L\'epine\altaffilmark{3}, R. Michael
  Rich\altaffilmark{4}, Michael M. Shara\altaffilmark{3}, Kelle L.
  Cruz\altaffilmark{3,5}, and Andrew Skemer\altaffilmark{6}}

\altaffiltext{1}{Based on observations performed with the Laser Guide
  Star Adaptive Optics system at the Lick Observatory, operated by the
  University of California system.}
  
\altaffiltext{2}{Based on observations conducted at the MDM
  observatory, operated jointly by the University of Michigan,
  Dartmouth College, the Ohio State University, Columbia University,
  and the University of Ohio.}

\altaffiltext{3}{Department of Astrophysics, Division of Physical Sciences,
American Museum of Natural History, Central Park West at 79th Street,
New York, NY 10024, USA}

\altaffiltext{4}{Department of Astrophysics, University of California
  at Los Angeles, Los Angeles, CA 90095, USA, rmr@astro.ucla.edu}

\altaffiltext{5}{NSF Astronomy and Astrophysics Postdoctoral Fellow}

\altaffiltext{6}{Department of Astronomy and Steward Observatory, 
University of Arizona, 933 N Cherry Ave., Tucson, AZ 85721, 
USA, askemer@as.arizona.edu}

\begin{abstract}
We report the discovery of a companion to the high proper motion star
LHS 1589, a nearby high-velocity, low-mass subdwarf. The companion
(LHS 1589B) is located $0.224\pm0.004\arcsec$ to the southwest of the
primary (LHS 1589A), and is 0.5 magnitude fainter than the primary in
the $K_s$ band. The pair was resolved with the IRCAL infrared camera
at Lick Observatory, operating with the Laser Guide Star Adaptive
Optics system. A low-resolution spectrum of the unresolved pair
obtained at the MDM observatory shows the source to be consistent
with a cool subdwarf of spectral subtype sdK7.5. A photometric
distance estimate places the metal-poor system at a distance
$d=81\pm18$ parsecs from the Sun. We also measure a radial velocity
$V_{rad}=67\pm8$km s$^{-1}$ which, together with the proper motion
and estimated distance, suggests that the pair is roaming the inner
Galactic halo on a highly eccentric orbit. With a projected orbital
separation $s=18.1\pm4.8$ AU, and a crude estimate of the system's
total mass, we estimate the orbital period of the system to be in the
range 75 yr $< P <$ 500 yr. This suggests that the dynamical mass of
the system could be derived astrometrically, after monitoring the
orbital motion over a decade or so. The LHS 1589AB system could thus
provide a much needed constraint to the mass-luminosity relationship
of metal-poor, low-mass stars.
\end{abstract}

\keywords{binaries: close --- techniques: high angular resolution ---
  stars: low-mass, brown dwarfs --- stars: subdwarfs --- stars:
  Population II --- Galaxy: solar neighborhood}

\section{Introduction}

Astrometric binaries, defined as physical pairs whose projected
orbital motion can be mapped out, are objects of high interest in
astrophysics because the gravitational masses of the components can be
directly calculated. Such systems are critical in constraining the
mass-luminosity relationship (MLR), one of the most fundamental
relationships in astrophysics
\citep{MPS97,Hetal99,Setal00,DFSBUPM00,Metal07}. The MLR allows one to
convert the luminosity function (LF), an observable quantity, into the
mass function or initial mass function (IMF) which is fundamental in
understanding stellar populations \citep{D98}. The MLR is particularly
critical to obtain reliable estimates of the stellar mass function and
baryonic content of the Galaxy \citep{CM97,PMFW00}. The MLR further
provides constraints to stellar evolutionary models \citep{BCAH98}. 

Differences have been observed in the luminosity functions of open and
globular clusters which are best explained by a dependence of the MLR
on metallicity \citep{HGTRJ96}. In low-mass stars, particularly the
M dwarfs/subdwarfs, the spectral energy distribution, dominated by
molecular bands of metal oxides and hydrides, is strongly affected by
variations in metallicity \citep{AH95}. A high metal content yields
stronger molecular bands, which shift the energy output from the
optical to the infra-red, making the star redder. Assuming the
bolometric luminosity to be simply dependent on stellar mass, the MLR
for metal-poor stars is thus expected to be shifted to brighter
magnitudes in the optical, and possibly also to fainter magnitudes in
the infrared. Theoretical models combined with observational evidence
suggest that metallicity effects on the MLR are more pronounced in the
visual bands than in the IR \citep{DFSBUPM00}, at least for M dwarfs
of near-solar metallicity ($-0.5<log[Fe/H]<0.0$). It has been
suggested that the MLR can be assumed to be independent of metallicity
in the K band \citep{Betal05}, although this remains to be
demonstrated for low-mass stars over a broad range of Fe/H values. It
is thus critical to verify the MLR calibration for stars over a wider
range of metallicities, particularly for very metal-poor
($log[Fe/H]<-1.0$) M subdwarfs.
 
There is extensive literature on the topic of low-mass astrometric
doubles and the calibration of the MLR for low-mass dwarfs of
near-solar metallicity (Population I). Early calibrations, in the
classic papers by \citet{LP87} and \citet{HM93}, relied on relatively
few systems with well determined orbits and masses. Other known double
stars have since had their orbits monitored with speckle
interferometry \citep{MDH99,Wetal00}, or with the Fine Guidance Sensor
on HST \citep{Fetal98,Hetal99,Tetal99,Betal00,Betal01}, yielding
accurate masses for both components. New close doubles have also been
resolved among nearby disk dwarfs with the NICMOS camera on HST
\citep{PSHB04}, with adaptive optics (AO) observations from CFHT
\citep{DFBUMP99}, with AO masking interferometry from Palomar
\citep{Letal06}, and with Laser Guide Star Adaptive Optics (LGSAO)
observations from Keck \citep{Petal06}.

\begin{figure*}
\epsscale{1.05}
\plotone{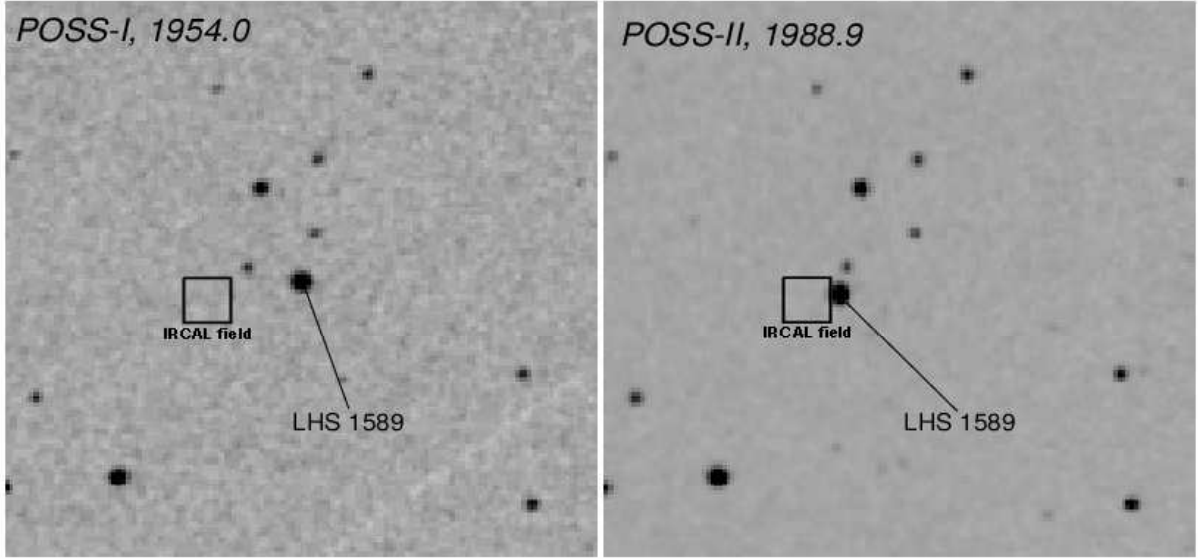}
\caption{Multi-epoch chart of the high proper motion star LHS 1589
  generated from the Digitized Sky Surveys. The field is 4.25$\arcmin$
  on the side. Left: first Palomar Sky Survey image (POSS-I) circa
  1954.0. Right: Second Palomar Sky Survey image (POSS-II) circa
  1988.9. The motion of LHS 1589 is quite apparent. The box shows the
  location of the field imaged with the Lick Adaptive Optics system on
  October 2006.}
\end{figure*}

In contrast, there exists no calibration of the mass-luminosity
relationship for the more metal-poor, low-mass subdwarfs
(Population II). Perhaps the chief reason for this is that most
metal-poor stars in the vicinity of the Sun are kinematically
associated with the Galactic halo population, and are thus relatively
scarce. Most studies estimate there is one halo star in the Solar
Neighborhood for each 200-300 disk stars, which means that one expects
to find only $\sim$20 halo stars within 25 parsecs of the
Sun. The MLR for low-mass, Pop II subdwarfs has also been lacking
because of the difficulty in measuring accurate metallicities in
low-mass stars, although significant progress have recently been made
on that front \citep{Bean06,WW06}. 

In any case, the existence of wide binaries among the halo population
has been demonstrated. Positive identification of low-mass subdwarfs
in lists of common proper motion doubles \citep{CG04} suggests that
the binary frequency among old metal-poor stars may be as high as that
of the local disk population. Recently, a companion was resolved only
$0.15\arcsec$ from the nearby ($\approx100pc$) metal-poor sdG2
subdwarf Ross 530 \citep{LHM06}. If they could be found, subdwarf
doubles with components of lower masses (sdK/sdM), and with
separations short enough for the orbital motion to be astrometrically
monitored, would be of great value to astronomy.

We have recently undertaken a massive study of the local population of
metal-poor, low-mass stars, following the identification of thousands
of high-velocity subdwarfs in the new LSPM-north catalog of stars with
proper motions $\mu>0.15\arcsec$ yr$^{-1}$ \citep{LS05} and in its
upcoming southern-sky complement, the LSPM-south (L\'epine {\it et al.}
2007, in preparation). Here, we report the discovery of a
close companion to the high proper motion star LHS 1589, which our
spectroscopic observations confirm to be a metal-poor, cool
subdwarf. In \S2, we describe our Adaptive Optics observations,
performed at Lick Observatory with the Laser Guide Star system. Our
spectroscopic observations are detailed in \S3. We discuss in \S4 the
prospects for using this system to constrain the mass-luminosity
relationship in the metal-poor, low-mass star regime.

\section{Adaptive optics observations}

\begin{figure}
\epsscale{0.85}
\plotone{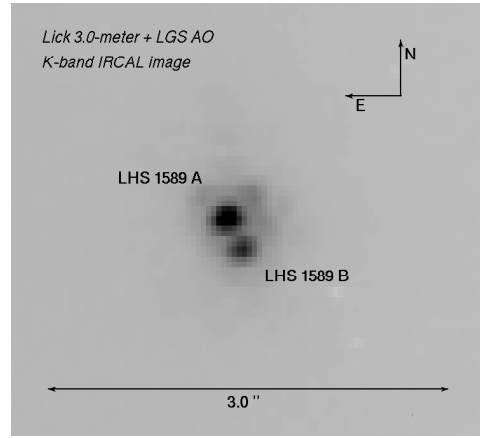}
\caption{Adaptive optics image of LHS 1589, resolving this high proper
  motion star into two distinct sources. The secondary is found
  $0.224\arcsec\pm0.004\arcsec$ to the southwest of the primary. The
  image shown here represents only the central part of the full
  IRCAL field, as was constructed from a set of five dithered frames.}
\end{figure}

From the LSPM-north catalog \citep{LS05}, we have assembled a list of probable low-mass subdwarfs with estimated (photometric) distances $d<100pc$ from
the Sun. A first subset of 18 targets
were observed on the nights of 2006 September 16-17, with
the Shane 3.0-meter telescope at Lick Observatory on Mount
Hamilton. Adaptive Optics (AO) observations were performed under
operation of the Laser Guide Star (LGS) system. High angular
resolution images were recorded with the IRCAL infrared camera, a
256$\times$256 pixels Rockwell PICNIC HgCdTe array \citep{L00}.
The camera has an effective field of view of approximately
$19.4\arcsec\times19.4\arcsec$ in the focal plane. It is
however slightly anamorphic, with a measured plate scale of 0.0754
$\arcsec$ pixel$^{-1}$ in the direction of Right Ascension, and 0.0780
$\arcsec$ pixel$^{-1}$ in the direction of Declination \citep{P06}. We
operated the laser guide star system using the science target as a
reference tip/tilt star. A total of 5 images were obtained in the K
band with a 5 point dithering pattern. 

No companions were detected in 17 of the 18 targets. These stars
yielded radially symmetric images, with the first Airy ring clearly
visible. We conclude that these stars are either single star systems,
or that they are multiples unresolved to within the diffraction limit
of the telescope.

One of our targets, the high proper motion star LHS 1589
(see Figure 1), was however resolved into a pair of close point
sources. All five images from the dithering pattern showed the same pair of resolved point sources, with similar
separation and orientation. A composite image was built using the {\em
  drizzle} algorithm in the IRAF STSDAS package, halving the effective
pixel scale to $0.0377\arcsec\times0.0390\arcsec$. Figure 2 shows the
central $3.5\arcsec\times3.5\arcsec$ part of the composite
image. Point spread function fitting of the two resolved sources
yields an angular separation $\rho=0.224\arcsec\pm0.004\arcsec$, with
a position angle on the sky from component A to B of
$\phi=206.3\pm1.3$ degrees.

We can rule out with near certainty the possibility that the two stars
are a chance alignment. Each component is bright enough to be detected
in any of the POSS-I or POSS-II images, but scans from the Digitized Sky Surveys (DSS) covering the piece of sky imaged by IRCAL is clearly devoid of any optical source
down to the magnitude limit of the photographic plates (see
Fig.1). Likewise, an image from the 2MASS data archive shows a single,
unresolved source near the location of LHS 1589, though the 2MASS data
was acquired in 1999, at a time when LHS 1589 was $>5\arcsec$ away from its
current location. Any background source located near the
current location of LHS 1589 would have been clearly resolved in the
2MASS image. Hence, the two point sources are most likely co-moving
stars. The only alternative would be for component B to be a strongly
variable or transient source that just happened to erupt within a
short angular distance of LHS 1589 at the time of the AO imaging, a
very unlikely event.

Component B currently lies to the South-West of the brighter component
A. Based on our aperture photometry, the secondary is 0.52 magnitudes
fainter that the primary in the K$_s$ band, which we can write down as
$\Delta K_s = (K_s)_B - (K_s)_A = 0.52$. Infra-red 2MASS photometry of
the unresolved source yields a total magnitude $(K_s)_{A+B}$=11.33
mag. We can derive magnitudes for the individual components using:
\begin{equation}
(K_s)_A =  (K_s)_{A+B} + 2.5 \log{ 1 + 10^{-\Delta K_s/2.5} },
\end{equation}
from which we calculate K$_s$=11.85 mag for component A, and 
$K_s$=12.36 mag for component B.
 
%
%
%
%

\section{Spectroscopic observations}

A red optical spectrum of the unresolved binary system was collected
at the MDM observatory with the 1.3m McGraw-Hill telescope on the
night of 14 October 2006. We used the Mark III spectrograph with the
300 lines/mm grating blazed at 8000\AA. The spectrum was recorded
with a LORAL $2048\times2048$ CCD camera (``Wilbur''), yielding a
spectral resolution of 3.11\AA\ per pixel. Reduction was done with
IRAF, including standard flat-fielding, sky-subtraction, extraction,
calibration, and telluric correction. Wavelength calibration was
determined from a NeArXe arc spectrum collected right after the
exposure on the star, yielding an estimated calibration accuracy of
$\pm0.35$\AA. Flux calibration was determined based on observations of
the spectrophotometric standard star Hiltner 600.

\begin{figure}
\epsscale{1.55}
\plotone{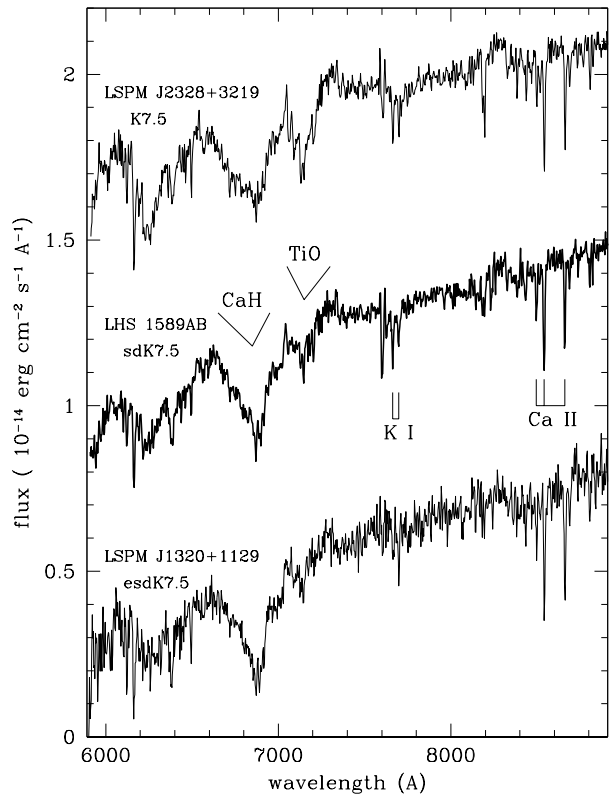}
\caption{Composite spectrum of the halo subdwarf binary LHS 1589AB. The
  strong CaH molecular band near 7000\AA\ flanked by a relatively weak 
  TiO bands indicates that the star is a metal-poor, low-mass
  subdwarf. Based on the strength of the CaH band, the spectral
  subtype of the composite spectrum is sdK7.5. Two other stars from
  our MDM spectroscopic survey are shown for comparison: the K7.5
  dwarf LSPM J2328+3219, and the esdK7.5 extreme subdwarf LSPM
  J1320+1129.}
\end{figure}

The reduced, composite spectrum is displayed in Figure 3. The
relatively weak TiO/CaH band ratio is consistent a cool subdwarf of
spectral type sdK-sdM. We calculated the three spectroscopic indices
CaH2, CaH3, and TiO5, which measure the depth of the CaH and TiO
molecular bandheads near 7040\AA\ \citep{RHG95}, finding CaH2=0.781,
CaH3=0.890, TiO5=0.906. In the revised classification system of
\citet{LRS07}, these values yield a spectral type  sdK7.5. The TiO/CaH
bandstrength ratio, which is the main metallicity diagnostic in cool
subdwarfs, is quite weak in LHS 1589AB, and the star actually falls
just short of being classified an extreme subdwarf (esdK7.5).

A direct comparison with spectra from other cool dwarfs and subdwarfs
confirms this impression. Figure 3 shows MDM spectra of the high
proper motion stars LSPM J2328+3219 and LSPM J1320+1129, respectively
classified as K7.5 and esdK7.5, respectively. Each is typical of stars
of its own metallicity subclass, and both have CaH bandstrengths
similar to those observed in LHS 1589AB. The TiO bands in LHS 1589
are clearly intermediate between the two comparison objects, but are
closer to those of the esdK7.5 star, which again indicates that LHS
1589AB consists of two subdwarfs (sd) close to the extreme subdwarf
(esd) metallicity class.

A radial velocity was calculated from the shift of individual lines in
the K II doublet and Ca II triplet, whose centroids were calculated in
IRAF with the SPECRED package. With a 1-pixel resolution of 3.11\AA\
and signal-to-noise ratio of 30, we estimate that we can measure
radial velocities to $\approx8 km/s$ . After correction for the
Earth's orbital motion, we determine a heliocentric radial velocity
$V_{rad}=+67\pm8$km/s.

\section{Discussion}

The star LHS 1589 was first discovered as a high proper motion star in
the Lowell Proper motion survey \citep{GBT61}, and was initially known
as G 79-69. The star was later included in the Luyten Half Second
catalog \citep{L79} under the catalog number 1589. Olin J. Eggen
also included the object is his catalog of stars with proper motions
$\mu>0.7\arcsec$ yr$^{-1}$ under the designation PM 03436+1134. Eggen
also recognized it to be a metal-poor subdwarf based on its relatively
blue $R-I$ color \citet{E83}. 

It is critical that the distance to the pair be properly measured so
that we can estimate its projected physical separation and probable
orbital period. This in turn will determine the potential for the
system to have its orbital motion mapped out astrometrically. We have
compiled published trigonometric parallaxes of spectroscopically
confirmed, nearby subdwarfs \citep{M92,VA95}. Figure 4 shows the
resulting $[M_{K_s},V-K_s]$  color-magnitude diagram. Optical V
magnitudes are from the original parallax tables, while infra-red
$K_s$ magnitudes are from the 2MASS All-Sky Point Source catalog
\citep{C03}. Spectroscopically confirmed subdwarfs (sdK,sdM) and
extreme subdwarfs (esdK, esdM) are plotted with distinct symbols in
Fig. 4 (filled triangles and open circles, respectively). The list
includes  subdwarfs classified by \citet{G97} and \citet{RG05}, as
well as subdwarfs classified in our own spectroscopic survey (L\'epine
et al. 2007, in preparation).

%
%
%
%
%
%
%

\begin{figure}
\epsscale{1.2}
\plotone{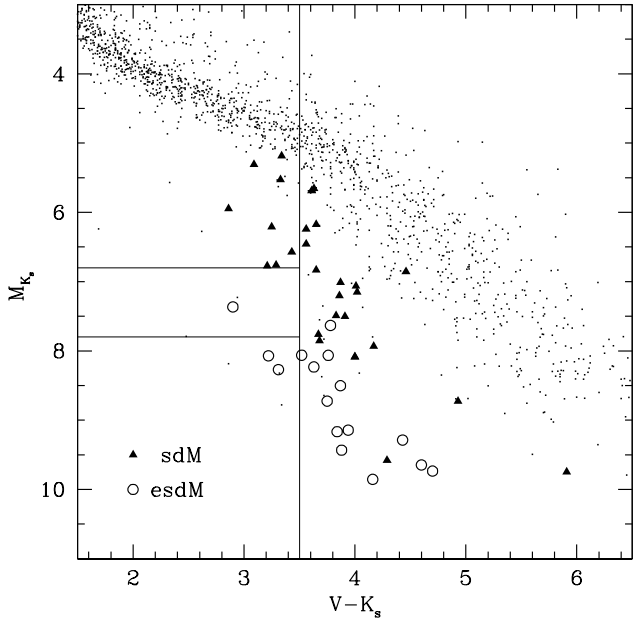}
\caption{Color-magnitude diagram of nearby stars with measured
  trigonometric parallaxes, from various sources. Spectroscopically
  confirmed subdwarfs (sdM) are shown as filled triangles;
  spectroscopically confirmed extreme subdwarfs (esdM) are shown as
  open circles. Component A in LHS 1589 has an estimated color
  V-K$_s$=3.5 (vertical line). Based on our spectroscopy (Fig.3), we
  assume that LHS 1589 is a subdwarf located close to the extreme
  subdwarfs, from which we estimate an absolute magnitude
  M$_{K_s}\approx7.3\pm0.5$ (horizontal lines). This in turns suggests
  a distance $d=81\pm18$pc for the system.}
\end{figure}

%
%
%

As can be seen in Fig.4, the color-magnitude relationship has a strong
dependence on metallicity for stars with $3<V-K_s<5$, with more
metal-poor stars being less luminous at a given color. The parallax
data suggest that $\Delta M_{K_s} \sim 1.7 \Delta (V-K_s)$ for sdK/sdM
subdwarfs. Our own measurements give $(K_s)_B - (K_s)_A=0.52$, from
which we estimate that $\Delta (V-K_s) = (V-K_s)_B - (V-K_s)_A$=0.31,
which yields $\Delta V = V_B-V_A = 0.83$. This suggests that component
B is about 2.15 times fainter in the V band than component A. The
integrated V band magnitude of the system is $V_{A+B}$=14.93, as
measured by \citet{HD80}. We can use Equation 1 again to calculate the
V magnitudes of components A and B, substituting $V$ for $K_s$ in the
equation. We find $V_A=15.35$ for the primary, and $V_B=16.18$ for the
secondary. This in turn suggests optical-to-infrared colors
$(V-K_s)_A=3.5$ and $(V-K_s)_B=3.8$. 

Figure 4 displays our best guess at the absolute magnitude of LHS
1589A.  If we assume $V-K_s\approx3.5$, and further assume the star to
be near the limit between sdK-sdM and esdK-esdM subdwarfs, then we
find that LHS 1589A has an absolute magnitude $M_{K_s}=7.3\pm0.5$. An
independent estimate of the absolute magnitude can be obtained based
on the star's spectrum. Relationships between the spectral subtype and
absolute magnitude have been calibrated by \citet{LRS03} and
\citet{RG05} for groups of dwarfs, subdwarfs, and extreme
subdwarfs. For an sdK7.5 subdwarf, just halfway between the subdwarf
and extreme subdwarf classes, the subtype-luminosity relationships are
consistent with an absolute magnitude $M_{K_s}\approx6.9$, which
agrees quite well with the color-magnitude estimate given above.

Based on an absolute magnitude $M_{K_s}=7.3\pm0.5$, we derive a
photometric distance $d=81\pm18$pc. The only recorded parallax for LHS
1589, a relatively crude $2.1\pm7.9$ milliarcsecond \citep{HD80}, is
consistent with our photometric distance estimate at the
$\approx1.3\sigma$ level. At a distance of 81pc, and assuming
$V_{rad}$=+67 km s$^{-1}$, the system would have components of motion
$(U,V,W)=(-155,-240,+117)$ km s$^{-1}$ in the local standard of
rest. An integration in the Galactic potential yields a very eccentric
orbit sharply inclined to the Galactic plane, consistent with a
membership with the inner Galactic Halo.


\begin{deluxetable}{lr}
\tabletypesize{\scriptsize}
\tablecolumns{4} 
\tablewidth{180pt} 
\tablecaption{Summary of data on LHS 1589AB} 
\tablehead{
\colhead{Observable} & 
\colhead{value}
}
\startdata
Primary R.A. (J2000)\tablenotemark{a}  & 03 46 24.53 \\
Primary Decl. (J2000)\tablenotemark{a} & +11 43 12.0 \\
$\mu$R.A.($\arcsec$ yr$^{-1}$)  &  0.783 \\
$\mu$Decl.($\arcsec$ yr$^{-1}$) & -0.178 \\
V$_{A+B}$ (mag) & 14.93 \\
J$_{A+B}$ (mag) & 12.06 \\
H$_{A+B}$ (mag) & 11.51 \\
(K$_s$)$_{A+B}$ (mag) & 11.33 \\
(K$_s$)$_{B}$-(K$_s$)$_{A}$ (mag) & 0.52\\
composite spectral type & sdK7.5 \\
V$_{rad}$ (km s$^{-1}$) & +67$\pm$8 \\
Angular separation ($\arcsec$) & $0.224\pm0.004$\\
position angle (degrees) &  $206.3\pm1.3$ \\
\hline \\
\colhead{Derived quantity} & 
\colhead{value} \\
\hline \\
V$_A$ (mag) & 15.35 \\
(K$_s$)$_A$ (mag) & 11.85  \\
V$_B$ (mag) & 16.18 \\
(K$_s$)$_B$ (mag) & 12.36 \\
photometric distance (pc) & 81$\pm$18\\
U (km s$^{-1}$) & -155$\pm$8\\
V (km s$^{-1}$) & -240$\pm$16\\
W (km s$^{-1}$) & +116$\pm$9\\
m$_A$ (M$_{\odot}$) & $\approx$0.25\\
m$_B$ (M$_{\odot}$) & $\approx$0.30\\
 \enddata
\tablenotetext{a}{Positions are given at the 2000.0 epoch and for the
  J2000 equinox.}
\end{deluxetable}


Is it also possible to estimate the mass of the system, with the
caveat that existing mass-luminosity and mass-color relationships have
all been calibrated using dwarf stars of roughly solar metallicity,
while the components of LHS 1589 are really metal-poor subdwarfs. Both
the mass-luminosity relationship in the V band and the mass-color
($V-K_s$) relationship are significantly dependent on metallicity
\citep{Betal05,DFSBUPM00}, and thus cannot be used. The
mass-luminosity relationship in the infrared K band, however, does
appear to be nearly independent on metallicity. According to the
\citet{DFSBUPM00} calibration, LHS1589A, with $M_K=7.3\pm0.5$ should
have a mass in the range $0.2 M_{\odot} < m_A < 0.3
M_{\odot}$. Component B, with $M_K=7.8\pm0.5$, should have a mass in
the range $0.15 M_{\odot} < m_B < 0.25 M_{\odot}$. The total mass of
the system should thus be roughly in the range of $0.35 M_{\odot} <
m_{A+B} < 0.55 M_{\odot}$.

At first glance, these mass estimates seem suspiciously low. This is
because mass to spectral-subtype relationships for dwarf stars
indicate that a star of spectral subtype K7.5 should have a mass of
$\approx0.6$ M$_{\odot}$ \citep{BC96,H04}. If one assumes that the
mass to spectral-subtype relationship applies similarly to the
subdwarfs, then this would suggest a mass $\approx0.6$ M$_{\odot}$ for
LHS 1589A, clearly at odds with our the 0.2-0.3$ M_{\odot}$
estimate. However, one should note that this mass to spectral-subtype
relationship has only been calibrated for K and M dwarfs of roughly
solar metallicity, and has not been verified for sdK/sdM subdwarfs. In
fact, spectral subtypes are usually a measure of the effective
temperature, and subdwarfs are expected to be both smaller and hotter
than dwarfs of comparable masses. This suggests that a subdwarf of a
given subtype is indeed likely to be less massive that a dwarf of
similar subtype. In fact, if we assume that $M_{K_s}$ is dependent
only on the mass of the star and not its metallicity, as suggested by
\citet{DFSBUPM00}, then the subtype-luminosity relationships of
\citep{LRS03} and \citep{RG05} indicate that an sdK7.5 star like LHS
1589A should have roughly the same mass  as an M3-M4 dwarf, which
itself has a mass of $\approx0.3$ M$_{\odot}$, a value consistent with
our mass estimate. 

It is also possible that the mass-luminosity relationship in the K band 
might not be as tight as suggested in \citet{DFSBUPM00}. It
would make sense for subdwarfs to be slightly underluminous
in the K band, to compensate for the fact that they are overluminous
in the optical, and assuming that stars of a given mass all have the
same bolometric luminosities. In which case, the current
mass-M$_{K_s}$ relationships, calibrated on K-M dwarfs, would indeed
underestimate the mass of a subdwarf of the same M$_{K_s}$. The only
way to resolve this conundrum is to obtain dynamical masses for
components of subdwarf binaries.

Assuming for now that the total mass of the system is in the range
0.35 M$_{\odot} < m_{A+B} < 0.55 M_{\odot}$, then with a projected
orbital separation $s=18.1\pm4.8$ AU we estimate an orbital period in
the range 75 yr $ \lesssim P \lesssim$ 500 yr. At the lower orbital
period regime, a fraction of the orbit could be mapped within a
decade, from which the total dynamical mass of the system could be
constrained with much greater accuracy \citep{Setal06}.

\section{Conclusions}

The low-mass subdwarf double LHS 1589AB should be
considered a high priority target for high angular resolution,
astrometric monitoring. The system is a good candidate for a direct
measurement of dynamical masses, and may prove critical for
calibrating the mass-magnitude relationship in metal-poor stars. At
the very least, an accurate parallax for the system should be
obtained, in order to estimate how much of an investment in time would
be required to map out the orbital motion. Systems with much shorter
orbital separations would clearly be more useful, however. Our
demonstration that astrometric doubles do exist {\em and} can be
detected among nearby subdwarfs, provides renewed impetus in the
search for additional systems. We are currently expanding our LGS-AO
survey with the aim of surveying all known low-mass subdwarfs within
100 parsecs of the Sun.

%
%
%
%
%
%
%
%
%
%
%
%

\acknowledgments

{\bf Acknowledgments}

The authors wish to thank the staff at Lick Observatory, in particular
Elinor Gates, for their operation of the Laser Guide Star system and
for outstanding support. This research program was supported by NSF
grant AST-0607757 at the American Museum of Natural History. SL and
MMS also gratefully acknowledge support from Mr. Hilary Lipsitz.
KLC is supported by an NSF Astronomy and Astrophysics Postdoctoral
Fellowship under award AST-040148.



\begin{thebibliography}{}

\bibitem[Allard \& Hauschildt(1995)]{AH95}
Allard, F., \& Hauschildt, P. H. 1995, \apj, 445, 433

\bibitem[Baraffe \& Chabrier(1996)]{BC96}
Baraffe, I., \& Chabrier, G. 1996, \apj, 461, L51

\bibitem[Baraffe {\it et al.}(1998)]{BCAH98}
Baraffe, I., {\it et al.} 1998, \aap, 337, 403

\bibitem[Bean {\it et al.}(2006)]{Bean06}
Bean, J. L. {\it et al.} 2006, \apj, 652, 1604

\bibitem[Benedict {\it et al.}(2000)]{Betal00}
Benedict, F., {\it et al.} 2000, \aj, 120, 1106

\bibitem[Benedict {\it et al.}(2001)]{Betal01}
Benedict, F., {\it et al.} 2001, \aj, 121, 1607

\bibitem[Bonfils {\it et al.}(2005)]{Betal05}
Bonfils, X., {\it et al.} 2005, \aap, 442, 635

\bibitem[Chabrier \& M\'era(1997)]{CM97}
Chabrier, G., \& M\'era, D. 1997, \aap, 328, 83

\bibitem[Chanam\'e \& Gould(2004)]{CG04}
Chanam\'e, J., \& Gould, A. 2004, 601, 289

\bibitem[Cutri {\it et al.}(2003)]{C03}
Cutri, R. M. {\it et al.} 2003, The 2MASS All-Sky Catalog of Point
Sources, University of Massachusetts and Infrared Processing and
Analysis Center, (IPAC/California Institute of Technology)

\bibitem[D'Antona(1998)]{D98}
D'Antona, F. 1998, The Stellar Initial Mass Function, APS Conference Series, 
Vol. 142, p. 157

\bibitem[Delfosse {\it et al.}(1999b)]{DFBUMP99}
Delfosse, X. {\it et al.} 1999, \aap, 344, 897

\bibitem[Delfosse {\it et al.}(2000)]{DFSBUPM00}
Delfosse, X. {\it et al.} 2000, \aap, 364, 217

\bibitem[Eggen(1983)]{E83}
Eggen, O. J. 1983, \apjs, 51, 183

\bibitem[Franz {\it et al.}(1998)]{Fetal98}
Franz, O. G., {\it et al.}, 1998, \aj, 116, 1432

\bibitem[Giglas, Burnham, \& Thomas(1961)]{GBT61}
Giclas H. L., Burnham, R., \& Thomas N.G. 1961, Lowell Obs. Bull., 5, 61

\bibitem[Gizis(1997)]{G97}
Gizis, J. E. 1997, \aj, 113, 806 (G97)

\bibitem[Golimowski {\it et al.}(2000)]{Getal00}
Golimowski, D. A., {\it et al.} 2000, \aj, 120, 2082

\bibitem[Harrington \& Dahn(1980)]{HD80}
Harrigton, R. S., \& Dahn, C. C. 1980, \aj 85, 454

\bibitem[Henry \& McCarthy(1993)]{HM93}
Henry, T. J., \& McCarthy, D. W. 1993, \aj, 106, 773

\bibitem[Henry {\it et al.}(1999)]{Hetal99}
Henry, T. J., {\it et al.} 1999, \apj, 512, 864

\bibitem[Henry(2004)]{H04}
Henry, T. J. 2004, ASP Conf. Series, 318, 159

\bibitem[von Hippel {\it et al.}(1996)]{HGTRJ96}
von Hippel, T., {\it et al.} 1996, \aj, 112, 192

\bibitem[Law {\it et al.}(2006)]{LHM06}
Law, N. M., Hodgkins, S. T., \& Mackay, C. D. 2006, \mnras, 368, 1917

\bibitem[L\'epine, Rich, \& Shara(2003)]{LRS03}
L\'epine, S., Rich, R. M., \& Shara, M. M. 2003, \aj, {\it submitted}

\bibitem[L\'epine \& Shara(2005)]{LS05}
L\'epine, S., \& Shara, M. M. 2005, \aj, 129, 1483

\bibitem[L\'epine, Rich, \& Shara(2007)]{LRS07}
L\'epine, S., Rich, R. M., \& Shara, M. M. 2007, \apj, {\it submitted}

\bibitem[Liebert \& Probst(1987)]{LP87}
Liebert, J., \& Probst, R. G. 1987, \araa, 25, 473

\bibitem[Lloyd {\it et al.}(2000)]{L00}
Lloyd, J. P. {\it et al.} 2000, Proc. SPIE, 4008, 814

\bibitem[Luyten(1979)]{L79}
Luyten W. J. 1979, LHS Catalogue: a catalogue of stars
with proper motions exceeding 0.5" annually, University of Minnesota,
Minneapolis ({\it CDS-ViZier catalog number I/87B})

\bibitem[Malkov {\it et al.}(1997)]{MPS97}
Malkov, O. Yu., Piskunov, A. E., \& Shpil'Kina, D. A. 1997, \aap, 320,
79

\bibitem[Martinache {\it et al.}(2007)]{Metal07}
Martinache, F. {\it et al.} 2007, \apj, 661, 496

\bibitem[Mason {\it et al.}(1999)]{MDH99}
Mason, B. D., Douglas, G., \& Hartkopf, W. I. 1999, \aj, 117, 1023

\bibitem[Monet {\it et al.}(1992)]{M92}
Monet, D. G., {\it et al.} 1992, \aj, 103, 638

\bibitem[Perrin(2006)]{P06}
Perrin, M. D. 2006, Ph.D. Thesis, UC Berkeley

\bibitem[Phleps {\it et al.}(2000)]{PMFW00}
Phleps, S., {\it et al.} 2000, \aap, 356, 108

\bibitem[Pravdo {\it et al.}(2004)]{PSHB04}
Pravdo, S. H., {\it et al.} 2004, \aj, 617, 1323

\bibitem[Lloyd {\it et al.}(2006)]{Letal06}
Lloyd, J. P. {\it et al.} 2006, \apj, 650, 131

\bibitem[Pravdo {\it et al.}(2006)]{Petal06}
Pravdo, S., {\it et al.} 2006, \apj, 649, 389

\bibitem[Reid, Hawley, \& Gizis(1995)]{RHG95}
Reid, I. N., Hawley, S., \& Gizis, J. E. 1995, \aj, 110, 1838

\bibitem[Reid \& Gizis(2005)]{RG05}
Reid, I. N., \& Gizis, J. E. 2005, \pasp, 117, 676

\bibitem[Schaefer {\it et al.}(2006)]{Setal06}
Schaefer, G. H., {\it et al.} 2006, \aj, 132, 2618

\bibitem[S\'egransan, D. {\it et al.}(2000)]{Setal00}
S\'egransan, D. {\it et al.} 2000, \aap, 364, 665

\bibitem[Torres {\it et al.}(1999)]{Tetal99}
Torres, G., {\it et al.} 1999, \aj, 117, 562

\bibitem[van Altena {\it et al.}(1995)]{VA95}
Van Altena W.F., Lee J.T., \& Hoffleit E.D. 1995, The General
Catalogue of Trigonometric Stellar Parallaxes, Fourth Edition, Yale
University Observatory

\bibitem[Woitas {\it et al.}(2000)]{Wetal00}
Woitas, J., {\it et al.} 2000, \aap, 353, 253

\bibitem[Woolf \& Wallerstein(2006)]{WW06}
Woolf, V. M., \& Wallerstein, G. 2006, \pasp, 118, 218

\end{thebibliography}
\end{document}